\documentclass[final,review,5p,times,twocolumn]{elsarticle}

\usepackage{graphicx}
\usepackage{bm}
\usepackage{showkeys}
\usepackage{url}

\journal{Physica D}

\begin{document}

\begin{frontmatter}


\title{Relating statistics to dynamics
in axisymmetric homogeneous turbulence }

\author[label1]{Fabien S. Godeferd\corref{fabien.godeferd@ec-lyon.fr}}
\address[label1]{LMFA CNRS UMR 5509, \'Ecole Centrale de Lyon, Universit\'e de Lyon, France\\
  36 avenue Guy de Collongue F-69131  \'Ecully\\
\indent E-mail:~~{\tt fabien.godeferd@ec-lyon.fr}}


\begin{abstract}
The structure and the dynamics of homogeneous turbulence are modified by the presence of body forces
such that the Coriolis or the buoyancy forces, which may render a wide
range of turbulence scales anisotropic. The corresponding statistical characterization of
such effects is done in physical space using structure functions, as well as in spectral space 
with spectra of two-point correlations, providing two complementary viewpoints.
In this framework, second-order and third-order structure functions
are put in parallel with spectra of two-point second- and third-order velocity correlation
functions, using passage relations. Such relations apply in the isotropic case, or for
isotropically averaged statistics, which, however, do not reflect the actual more complex
structure of anisotropic turbulence submitted to rotation or stratification.  This complexity
is demonstrated in this paper by orientation-dependent energy and energy transfer
spectra produced in both cases by means of a two-point statistical model for axisymmetric
turbulence. We show that, to date, the anisotropic formalism used in
the spectral transfer statistics is especially well-suited to analyze
the refined dynamics of anisotropic homogeneous turbulence, and that it can help
in the analysis of isotropically computed third-order structure function statistics often
used to characterize anisotropic contexts. 

\end{abstract}


\end{frontmatter}

\section{Introduction}

Homogeneous turbulence submitted to distortions such as solid body rotation,
stratification, or the Lorentz force in the MHD context, exhibit axisymmetric statistics, 
as a clear departure from isotropy. These anisotropic effects that arise 
due to  modified dynamics or energy exchange do not fall within the classical
description of turbulence in Kolmogorov's theory. 
The following general questions, pertaining to the understanding of anisotropic homogeneous
turbulence, can therefore be raised:
\begin{enumerate}
\item How can we characterize the
anisotropy of the flow, at one or two point, in physical or spectral statistical descriptions? 
\item Are  passage relations available between: (a) the two-point 
statistics of structure functions, necessarily
measured in physical space in experiments, and (b) kinetic energy and transfer spectra? 
(Which can be obtained
for high Reynolds number turbulence from statistical models.)
\item How can we
interpret the modified dynamics of anisotropic turbulence? (It can
be observed
  in terms of anomalous scalings of
energy spectra or  energy
transfer spectra  readily available in two-point statistical models.)
\item Can we exploit some features 
easily accessed in spectral space to refine the phenomenological analysis 
of the dynamics of turbulence in physical space?  
\item Do Kolmogorov scalings available for high Reynolds number isotropic turbulence
apply to isotropically integrated statistics of anisotropic turbulence?
\item What level of complexity do we have to introduce to go beyond the
mere isotropic description, \textit{e.g.} starting with axisymmetric flows?
\end{enumerate}   

As a bootstrap, we start, hereafter in this introductory section, by reviewing a few of the
existing works related to these issues, and introduce some of the
existing results and formalism available in the statistical characterization
of axisymmetric turbulence.
In section~\ref{sec1}, we describe the relationships
available for isotropic turbulence for second- and third-order statistics
appearing in both Lin's and K\'arm\'an-Howarth's equations,  and suggest to extend
the description of the velocity increment statistics to the anisotropic
case, and to relate them to the modified dynamics in axisymmetric homogeneous turbulence, 
especially energy and transfer spectra. Results for the latter, in the case of
stably stratified or rotating turbulence, are provided in section~\ref{edqnmres}. We
choose to use an anisotropic two-point statistical model, described
shortly in section~\ref{secedqnm}, since it permits to reach higher  
Reynolds number and smoother statistics than direct numerical simulation.
Assuming isotropized passage relations, we then derive second- and
third-order velocity structure functions, presented in section~\ref{physres},
and discuss the results against existing scalings for isotropic turbulence.

The spectral formalism for homogeneous
turbulence allows to remove altogether the role of pressure, and hence permits
a highly refined analysis of the dynamics of anisotropic turbulence, and
on its sources; for instance
from the role of inertial waves interactions in rotating turbulence, or for the 
dynamics of stably stratified turbulence in which the transfers are split
between  internal waves and potential vorticity interactions.
Even in flows which are thought to be isotropic, in experimental or numerical
realizations, a degree of anisotropy may be hidden, depending on how
the characterization of the isotropy of the flow has been done:
the isotropy of
\textit{rms} velocity components or of integral length scales only characterizes
the large scales, whereas vorticity or dissipation can be used for the
smaller ones. The computation of anisotropic spectra
allows to quantify such scale-dependent level of isotropy, or anisotropy,
and to isolate the most important energy transfer contributions.
Alternately, the statistics in physical space 
do not easily permit the removal of pressure from balance equations, and
thus prevent from a tractable access to modal decompositions (\textit{e.g.} the
toroidal/poloidal decomposition). It however applies to locally inhomogeneous
flows, which is not easily accessible to spectral analysis. 
We therefore
emphasize the importance of both spectral and physical viewpoints, and
the fact that they are \textit{complementary}, which we expose and discuss in this paper,
taking rotating or stably stratified turbulence as supporting evidences.

In what follows in this section, we review a few studies which have
been devoted to the characterization of anisotropy in turbulence, 
from the point of view of directional statistics in physical space --- including
both anisotropy of velocity components or dependence on the separation vector ---  
and studies about the anisotropic spectral scalings in anisotropic turbulence,
related to the dynamics of the flow.

If  initially isotropic, homogeneous turbulence can be rendered anisotropic  by introducing
an external distortion on the flow: solid body rotation is present for instance
in geophysical flows and acts through the Coriolis force, 
as well as a buoyancy force in density- or temperature- stratified layers.
In conducting fluids, the action of an external magnetic field also modifies
the symmetries by means of the Lorentz force. In these three examples, the
intensity of the corresponding force depends on the orientation of the 
fluid motion with respect
to either the rotation axis, the gravity axis, the background magnetic field
axis. Wave propagation can also be present, \textit{e.g.} inertio-gravity waves or Alfv\'en
waves. They provide an anisotropic way of redistributing energy in terms of scale
and direction, such that the dynamics of energy exchange in the turbulence is
strongly modified. We focus here on the effects of solid body rotation and
stable density stratification on homogeneous turbulence.

Studies about anisotropic turbulence were proposed and stemmed from 
various preoccupations. The theory of axisymmetric turbulence was
regarded as a logical extension to the theory of isotropic turbulence,
and was developed by Batchelor~\cite{batchelor46} --- in which parallel
and perpendicular  directions (with respect to the axis of symmetry)
were distinguished to 
express statistical quantities, \textit{e.g.}
the dissipation --- and by~Chandrasekhar \cite{chandrasekhar50} in the 50s. 
In these studies, the imposed symmetries
do not account for the possible presence of helicity, unlike the extension proposed by
Lindborg about the kinematics of axisymmetric turbulence \cite{lindborg95}.
More recently, some exact vectorial laws were also proposed by Galtier for rotating homogeneous
turbulence, exposing the need for a transverse/longitudinal  components
distinction (with respect to the separation vector)
 when computing velocity structure functions \cite{galtier09}.
An extensive discussion and review of the anisotropy in turbulent flows was proposed
by Biferale \& Procaccia \cite{biferale_anisotropy_2005} using the  
symmetry group analysis (SO(3) for Navier-Stokes equations). These authors
discuss Kolmogorov's theory and how to relate anisotropic flows to the
$n$-th order structure functions, especially looking at the 4/5 law for
the third-order one, which includes the isotropy hypothesis.

On the one hand, the structure-function approach has recently been used for 
an experimental characterization of rotating turbulence dynamics by
Lamriben \textit{et al.}
\cite{lamriben2011}, and for mesoscale turbulence in the atmosphere by Lindborg
\& Cho \cite{LindborgCho},
both works including a discussion of the level of anisotropy to account for, as
a departure to classical Kolmogorov scalings of velocity increments moments.
In an experimental study concerning a turbulent jet, Xu \& Antonia \cite{xu_antonia07}  emphasize the 
importance of discriminating between the longitudinal and transverse velocity 
components when expressing the
 structure function in axisymmetric turbulence, and Oyewola
\textit{et al.} use structure functions to describe the anisotropy of the small scales in a
turbulent boundary layer \cite{oyewola}.
In situ experiments by Kurien \textit{et al.} in the atmospheric
boundary layer also permitted to exhibit the different scalings of the structure
functions, distinguishing between longitudinal and transverse increments, by 
separating the lowest order anisotropy contributions thanks to the SO(3)
symmetry group. \cite{kurien2000}

On the other hand, several efforts were devoted to establishing 
the scalings that should apply to two-point velocity correlation spectra
depending on the perpendicular $k_\perp$ or the parallel $k_\parallel$ 
wavenumber components with respect to the axis of symmetry.
In the geophysical context, one may retrieve atmospheric spectra  as in rotating  strongly
stratified turbulence. At small scales (several kilometers), 
Kolmogorov  $k_\perp^{-5/3}$ scaling
is retrieved, whereas at large scales (several hundreds of kilometers), depending on the
velocity component considered the scaling is either $k_\perp^{-3}$  (zonal wind)
or $k_\perp^{-2}$ (meridional wind). \cite{lindborg06} In the context of conducting fluids, plasmas and astrophysics,
Galtier has also proposed several spectral scalings with spectra of the form
 $k_\perp^{-\alpha}k_\parallel^{-\beta}$ to account for anisotropy in magnetohydrodynamic
turbulence \cite{galtier2005}. However, the physical arguments available for discussing the anisotropy of conducting fluid submitted to the Lorentz force are not available for
disentangling the intricate nonlinear dynamics of rotating turbulence, and, to some
extent of stably stratified turbulence.

In most of the above-mentioned works, spectra are used to characterize the cascade
of energy and the dynamics of anisotropic turbulence, or structure functions compared
with the isotropic theory scalings. Very few works are devoted to the relationship
between the two formalisms, and the current paper is a first attempt at providing 
 \textit{quantitative} information on how anisotropic spectra dynamics can be used
to interpret physical space velocity increment statistics.

Thus, taking into account the modifications of the turbulence
structure and dynamics due to external effects can be done at different levels. 
The statistical description of the velocity field 
can be done at a two-point level in physical space by the second-order
velocity correlation tensor 
$R_{ij}(\bm{r})=\langle u_i(\bm{x}) u_j(\bm{x}+\bm{r}\rangle$
where $\bm{r}$ is the separation vector.  Fourier-transforming this
tensor yields the spectrum tensor 
$\Phi_{ij}(\bm{k})={1}/(8\pi)^3\int R_{ij}(\bm{r})\exp(-\mathrm{i}\bm{k}\cdot\bm{r})\mathrm{d}\bm{r}$,
whose trace is $\Phi_{ii}(\bm{k})=E(k)/2\pi k^2$ in isotropic turbulence, with $k=|\bm{k}|$ and $E(k)$
is the kinetic energy spectrum. Kolmogorov theory supports the scaling $E(k) \sim  k^{-5/3}$ in isotropic
turbulence at high Reynolds number.

However, external distortions render the turbulent flow statistics dependent
on the orientation of $\bm{r}$ in physical space, \textit{e.g.} for the two-point correlations
$R_{ij}(\bm{r})$. In isotropic turbulence, this tensor depends only
on two scalar function, namely the longitudinal two-point correlation function $f(r)$,
and the transverse one $g(r)$ \cite{batchelor53}, and is expanded over
the tensors $\delta_{ij}$ and $r_ir_j/r^2$. In axisymmetric turbulence,
which is the case we consider here---where the axis of symmetry $\bm{n}$ is borne
by the rotation vector in rotating turbulence or the axis of gravity in stably stratified
turbulence---,
$R_{ij}(\bm{r})$ depends on four scalars and requires the addition of
the two expansion tensors $r_in_j/r$ and $n_in_j$ \cite{batchelor46}.

The anisotropic dependence of the two-point correlation
tensor in turn translates
into a dependence of the corresponding spectra  $\Phi_{ij}(\bm{k})$ 
on the orientation $\theta$ of the wavevector $\bm{k}$ with respect to 
the axis of symmetry $\bm{n}$ (the latter is the same as in physical space).
The spectra can similarly also be decomposed in a general way over four
scalars (see \textit{e.g.} \cite{cambon97} and \cite{sagaut08}).

This correspondence between the physical space two-point statistical formalism
of turbulence and the spectral description not only applies to second-order correlation tensors
and spectra, but it also extends to third-order statistics, and moreover
to velocity structure functions. The latter are of great use in Kolmogorov's
description of isotropic turbulence, and are described \textit{e.g.}
in \cite{monin} and in \cite{frisch}. Considering the velocity
increment $\delta\bm{u}=\bm{u}(\bm{x}+\bm{r})-\bm{u}(\bm{x})$, in
the isotropic context one commonly computes the $n^{th}$-order structure
function $\langle(\delta u)^n\rangle$ based on the longitudinal projection $u$
of the velocity vector onto the separation vector $\bm{r}$. The brackets
indicate  averaging using statistical homogeneity. Obviously, when
turbulence is anisotropic, one should also distinguish different
components in the structure functions and different orientations
of $\bm{r}$, although most of the theoretical
results are available only in the isotropic context. 
One of them is 
the exact
Kolmogorov four-fifth law  wherein the third-order structure function 
$D_{LLL}(r)=\langle(\delta u)^3\rangle=-4/5\varepsilon r$, valid for
$r$ in the inertial subrange, where $\varepsilon$ is the averaged
dissipation \cite{k41a}, or, in a different form proposed by Antonia \textit{et al.}~\cite{antonia97}:
\begin{equation}
\langle\delta u (\delta u_j\delta u_j)\rangle=-\frac{4}{3}\varepsilon r .
\label{anton}
\end{equation}
where repeated indices imply summation.
%

In the following, we investigate how anisotropic turbulence can recover
the above scalings, using the two-point statistical EDQNM model. 
The question about the convergence of the statistics of the structure functions
to the 4/5 limit has been addressed before. Antonia \& Burattini studied this limit
at increasing Reynolds numbers \cite{antonia06}, and it appears that the convergence
is very slow, and may not be yet reached in the existing experiments. However, this
low-Reynolds number effects has to be separated from the anisotropic effect 
in non isotropic flows. The leading order small-scales K41 scalings may be recovered
in anisotropic flows, provided suitable filtering has been applied to the
data, as shown with DNS fields by Taylor \textit{et al.} \cite{taylorkurien}.
In the following, the interpretation of the scaling laws provided by EDQNM
results  therefore mixes  the isotropic small-scales turbulent
behavior with the anisotropic contributions, since no such separation as
in Taylor \textit{et al.} has been made.

\section{The link between spectral dynamics and velocity structure functions}
\label{sec1}
The above-mentioned  structure functions provide a scale-dependent statistical characterization
of turbulence with a two-point separation. It is clear that they are related
to the two-point correlation functions, \textit{and their spectral
counterparts}, such that second-order velocity structure functions
are related to two-point velocity correlation spectra, and third-order structure functions to
energy transfer spectra (triple velocity correlations at two points). 

In the anisotropic context, and
when dealing with statistically axisymmetric turbulence, the orientation $\theta$ of the
Fourier vector $\bm{k}$ with respect to the axis of symmetry borne by  
   $\bm{n}$, has to be taken into
account, so that an additional dependence has to be introduced in the energy spectrum: $E(k,\theta)$.
Integrating the latter over  $\theta\in [-\pi,\pi]$  of course provides an equivalent spherically
averaged $E(k)$.
At this stage, the question of the resulting inertial scaling of the spectrum is raised.

The kinetic energy spectrum $E(k)$ and the nonlinear energy transfer spectrum $T(k)$ appear in
the  Lin equation as
\begin{equation}
\partial_t E(k)+2\nu k^2 E(k)=T(k) \label{eqLin}
\end{equation}
in the isotropic case, where $\nu$ is the viscosity, such that the total
dissipation is $\varepsilon=2\nu\int k^2 E(k)\textrm{d}k$. 
In the rotating case for instance, phase-scrambling anisotropically changes
the shape of the energy exchange term $T(k,\theta)$ in an equation
equivalent to~(\ref{eqLin}) (see next equation \ref{eqneuf}). The transfer 
contains third-order correlation terms, and is thus related to the third-order 
statistics that are used to quantify nonlinearity such as the skewness of
velocity gradient.

Regarding second-order moments, one can obtain  second-order structure
functions starting with the two-point correlation spectra.
For instance, 
\begin{equation}
\langle \left( \delta u_i  \delta u_i \right) \rangle  = 2 \int^{\infty}_0 \left(1 -
\frac{\sin kr}{kr} \right) E(k) \mathrm{d}k\ .
\label{eqqdeux}
\end{equation}
where repeated indices are summed, contains information
from all the velocity components.

Equation~(\ref{eqLin}) therefore relates the evolution of second-order
statistics---the energy spectrum---to third-order ones---the transfer.
The analogous in physical space of~(\ref{eqLin}) is 
the K\'arm\'an-Howarth equation
\begin{equation}
\partial_t\left(u'^2 f\right)=\left(\partial_r+\frac{4}{r}\right)\left[R_{LL,L}(r,t)
+2\nu\partial_r\left(u'^2 f\right)\right] \label{eqKH}
\end{equation}
in which $f(r)$ is the longitudinal two-point correlation function;
 the longitudinal two-point third-order correlation function is
$R_{LL,L}(r,t)=\langle u_i(\bm{x},t) u_m(\bm{x},t) u_i(\bm{x}+\bm{r},t) \rangle {r_m}/{r}$,
$r=|\bm{r}|$, and $(3/2)u'^2$ is the total kinetic energy. Upon examining~(\ref{eqLin}) and~(\ref{eqKH})
together, it is clear that if the dynamics of turbulence
is anisotropic, one might expect not only a different equilibrium between the
dissipative and nonlinear related terms of these equations, and  
a directional dependence with $\bm{k}$ of the Lin equation
for $E(\bm{k})$ transposed into anisotropy with respect to $\mathbf{r}$ in the
K\'arm\'an-Howarth equation.
The derivation of the latter equation relies on assumption of isotropy, 
but  the general K\'arm\'an-Howarth-Monin equation, which contains
 the same terms that depend on the separation \textit{vector} $\mathbf{r}$
can be derived as (see Frisch \cite{frisch}, p. 78 for a derivation, 
and \textit{e.g. } \cite{lamriben2011}):
\begin{equation}
\frac{1}{2}\frac{\partial }{\partial t}
\langle\mathbf{u}(\mathbf{x})\mathbf{u}(\mathbf{x+r}\rangle
=\frac{1}{4}\mathbf{\nabla}\langle\delta\mathbf{u}(\mathbf{r})|\delta\mathbf{u}(\mathbf{r})|^2\rangle
+\nu \nabla^2  \langle\mathbf{u}(\mathbf{x})\mathbf{u}(\mathbf{x+r}\rangle
\end{equation}

The dependence of the velocity, 
or velocity increments $\delta\bm{u}$, on the
direction of $\bm{r}$ with respect to $\bm{n}$, 
is therefore an important parameter in axisymmetric turbulence statistics.
For isotropic turbulence, Kolmogorov's theory (1941) only considers a scalar 
 longitudinal 
increment $\delta u$, 
provides the scaling 
$\langle(\delta u)^n\rangle\sim \left(\varepsilon r\right)^{n/3}$, and allows
to
draw from equation~(\ref{eqKH}) a simplified 
 relationship between second- and third-order structure functions 
  $$
    \langle\left(\delta u\right)^3\rangle=-\frac{4}{5}\varepsilon r+6\nu\partial_r\langle(\delta u)^2\rangle.
  $$
which yields, at infinite Reynolds number, 
the famous four fifths law 
\begin{equation}
\langle\left(\delta u\right)^3\rangle=-({4}/{5})\varepsilon r ,
\label{eqdix}
\end{equation}
or equation~(\ref{anton}).  \cite{antonia97}

A useful relationship can be established between third-order spectral statistics \textit{i.e.} the
nonlinear energy transfer $T(k)$, and the third-order structure function :
\begin{equation}
\langle\delta u\left(\delta q\right)^2\rangle=4r\int_0^{\infty}g(kr)T(k)\mathrm{d}k \label{thstr}
\end{equation}
where $g(kr)=(\sin kr - kr\cos kr)/(kr)^3$ and $(\delta q)^2=(\delta u_i\delta u_i)$. It shows the direct link of the third-order
structure function on the dynamics of turbulence. 

Formulas similar to equations (\ref{eqqdeux}) and (\ref{thstr})
rigorously derived for anisotropic turbulence are still not available 
in the general case, but, in the rotating case, the isotropic relationships apply, 
since the Coriolis force is not a production term in the balance equations, 
and drops out altogether in the isotropized balance equations, 
both in  physical and spectral spaces.
In general anisotropic turbulence, the
analogous of~(\ref{thstr}) implies a vector dependence of the structure functions
on $\bm{r}$. It also shows that a dynamics modified  by an external distortion applied to homogeneous
turbulence translates immediately in a modification of the structure functions.
It is therefore interesting to discuss the applicability of Kolmogorov scalings in
flows that contain some anisotropy. Attempts at such scalings for 
rotating turbulence are proposed by Galtier \cite{galtier09}, or by Lindborg \&
Cho  using atmospheric data. \cite{LindborgCho}
In the axisymmetric case, an extensive work is
in progress~\cite{ANISO}.

\section{Statistical model for the spectral anisotropy of axisymmetric turbulence}
\label{secedqnm}
We study a simplified case which nonetheless captures some important
anisotropic physical mechanisms in \textit{e.g.} geophysical flows: stable
stratification and rotation, taken into account in the
Boussinesq system of equations for an incompressible fluid:
\def\vu{\bm{u}}
\def\vn{\bm{n}}
\def\vnab{\bm{\nabla}}
\begin{eqnarray}
\label{eq:realnavier1}
{\partial_t \vu} + \vu \cdot \vnab \vu - \nu \nabla^2 \vu &=&
- \vnab p
- 2 \Omega  \; \vn \times \vu + b \; \vn, \\
\label{eq:buoyancy}
{\partial_t b} + \vu \cdot \vnab b - \chi \nabla^2 b &=&
-N^2 \vn \cdot \vu,\\
\label{eq:incomp}
\vnab \cdot \vu &=& 0,
\end{eqnarray}
where $b$ is the buoyancy field,
associated with fluid  density fluctuations around background
density augmented by mean density gradient.
 $N$ is the Brunt-V\"ais\"al\"a (buoyancy) frequency
and $\Omega$ the rotational frequency. 
The corresponding
terms, Coriolis force and buoyancy force, act on the
velocity field linearly and compete against the
nonlinear advection term. In this study, the separate
effects of either one is considered  (although the
combination of both characterizes some atmospheric
or oceanic flows, with a ratio $\alpha=2\Omega/N\sim 1/10$).

In rotating or stratified turbulence, the dynamics
of the flow is superimposed  with wave-like motion that
transfers energy differently from the classical turbulent
dynamics. In the linear limit, at large $N$ or $\Omega$,
one recovers wave turbulence, as a soup of superimposed
inertial waves or internal gravity waves.
For instance, ``turbulence'' and ``wave'' like
dynamics may be defined by splitting the velocity field
in the eigenmodes of the linearized
system~(\ref{eq:realnavier1})--(\ref{eq:incomp}), so that
the total turbulent energy of a rotating and stratified
flow can be divided into a
``vortex'' mode and a ``wave'' mode.
The linear evolution of the wave mode is
governed by he dispersion relation
$\sigma(\bm{k})=N \sin \theta$ for internal
waves and $\sigma(\bm{k})=2\Omega \cos \theta$ for inertial waves,
which depend on $\theta$, the polar angle with the vertical, whence
a directional dependence of the turbulent motion.

In order to study relatively high Reynolds number turbulence,
we introduce a statistical model which describes the evolution
of the two-point correlation spectra, whose dynamical equations
derive from~(\ref{eq:realnavier1})--(\ref{eq:incomp}), introducing
a closure known as EDQNM. The Eddy-Damped Quasi-Normal Markovian
model for isotropic turbulence was studied by several 
authors (see \textit{e.g.}  Orszag \cite{ORSZAG71}) decades
ago, although it has recently regained interest for its
ability to reach very high Reynolds numbers 
as demonstrated by the spectrum we computed with EDQNM, shown on figure~\ref{fig5}. 
(Similar 
figures are presented in \cite{bos_dynamics_2006}.)

   The more advanced version developed for rotating or
stably stratified turbulence  
(denoted EDQNM$_2$ since it differs significantly from the model
for isotropic turbulence) 
reflects more accurately the wave dynamics and is capable of taking into
account  anisotropic features. In the limit of very strong
rotation, for instance, it becomes identical to the
 wave turbulence closures described by Zakharov \cite{zakharov_kolmogorov_1992},
although in this book,  Zakharov \textit{et al.} do not consider anisotropic dispersion
relations as for instance inertial waves. 
An extensive description of the EDQNM$_2$ closure model
is provided in \cite{godeferd2003, cambon97}, as well
as comparisons with Direct Numerical Simulations (DNS) 
and Large Eddy Simulations (LES) which show a very good agreement and
illustrate the capability of the model to accurately represent
anisotropic turbulence.
We find that, among other closures, only the EDQNM$_2$ model for closing
the spectral transfer terms (the equivalent to $T$ in equation~\ref{eqLin}) 
 was capable of predicting all the anisotropic
features observed in the DNS and LES results.
 The model equations for the generalized
transfer terms, \textit{e.g.} $T^e(\bm{k})$ for the directional
kinetic energy spectrum,  involve sums of eight contributions (according to
polarities of triadic interactions),  weighted
by the rotation-dependent factor (or Brunt-V\"ais\"al\"a frequency in the stratified
case):
\begin{equation}
T^e=\sum_{\epsilon,\epsilon',\epsilon''=\pm 1}
\int_{\bm{k}+\bm{p}+\bm{q}=0}
\frac{\mathbf{\mathsf{S^e_{(QN)}}}}{\vartheta_{kpq}^{\epsilon,\epsilon',\epsilon''}}
\mathrm{d}^3\bm{p}
\label{eqneuf}
\end{equation}

                                      
The numerator of the integrand takes into account
the quasi-normal expansion for non-isotropic turbulence and is closed in terms 
of quadratic combinations of the
two-point correlation spectra, 
whereas the denominator involves viscous and eddy-damping effects through
\begin{equation}
\vartheta_{kpq}^{\epsilon,\epsilon',\epsilon''}=
\vartheta_{kpq}(1+2\mathsf{i}\vartheta_{kpq}
\Omega(\epsilon k_{\parallel}/k+\epsilon' p_{\parallel}/p+\epsilon'' q_{\parallel}/q)) \label{eqdamp}
\end{equation}
for a triad of wavevectors $\mathbf{k}+\mathbf{p}+\mathbf{q}=0$ with
polarities $\{\epsilon,\epsilon',\epsilon''\}=\{\pm 1\}^3$,
thus accounting for the  explicit linear rotation effects on triple correlation through the phases. 
Thus, the classical timescale $\vartheta_{kpq}$ in the isotropic non-rotating case is
replaced by the preceding triadic complex timescales.
As already mentioned, the above expression is consistent 
with wave turbulence results~\cite{BENNEY-SAFFMAN,bellet_wave_2006}.
The isotropic EDQNM model consists only of the Lin equations (\ref{eqLin})
in which the transfer term is closed using double products of energy 
spectra $E(k)$, and where the only dependence variable is the wavenumber $k$.

\section{Second- and third-order structure functions derived from spectral statistics}
%
The EDQNM$_2$ model for rotation and stable stratification consists of equations
for the two-point correlation spectra similar to (\ref{eqLin}) in which the nonlinear transfer
spectra in the right-hand-side are replaced by their closure. In the model
for rotating turbulence, four coupled equations are solved for the 
energy density spectrum $e(\mathbf{k})$, the helicity $h(\mathbf{k})$
and the polarization spectra $Z(\mathbf{k})$, each of which is expressed at each discretized spectral point
in the polar-spherical representation of wavevectors $\bm{k}$, that is as functions
of discretized $(k,\theta)$:
\begin{eqnarray}
\left({\partial \over \partial t} + 2 \nu k^2\right)e &=& T^e \nonumber \\
\left({\partial \over \partial t} + 2 \nu k^2 + 4{\rm i}\Omega {k_3
\over k} \right) Z &=& T^z \label {eq12} \\
\left({\partial \over \partial t} + 2 \nu k^2\right)h & =& T^h \nonumber
\end{eqnarray}
The transfer  spectra are also discretized in the same way.

For stratified
turbulence, the density spectrum of rescaled 
potential energy $\Phi_3(\mathbf{k})$ and the cross-correlation spectrum for density and
velocity $\Psi(\mathbf{k})$ are involved
\begin{eqnarray}
\left(\frac{\partial}{\partial t} + 2\nu k^2\right) \Phi_1 &=& T^1 \nonumber \\
\left(\frac{\partial}{\partial t} + 2\nu k^2\right) \left(\Phi_2+\Phi_3\right) &=& T^{2}+T^{3} \label{eqstrat}\\
\left(\frac{\partial}{\partial t} + 2\nu k^2 -2\mathrm{i}N\sin\theta_k \right) 
        \left( \Phi_2-\Phi_3+\mathrm{i}\Psi_R \right) &=& T^2-T^3+IT^{\Psi_R} \nonumber
\end{eqnarray}
along with the kinetic energy spectrum decomposed in poloidal and toroidal
spectra $\Phi_2(\mathbf{k})$ and $\Phi_1(\mathbf{k})$.

Equations (\ref{eq12}) and (\ref{eqstrat})
are therefore generalized versions of the Lin equation (\ref{eqLin}) and are exact
equations, unless a closure such as equation (\ref{eqneuf}) is applied.
Moreover, in both systems (\ref{eq12}) and (\ref{eqstrat}), the decomposition of the spectra
is optimized to yield simple and meaningful equations: the $Z$ equation in (\ref{eq12})
contains the oscillatory part due to the effect of rotation, and the last equation
of (\ref{eqstrat}) represents oscillations due to the kinetic/potential energy
exchange, mediated by the flux $\Psi$ produced by buoyancy.

The complexity of these systems of equations  and anisotropic
damping (\ref{eqdamp}) provided by the EDQNM$_2$ model
needs to be contrasted with the simplicity of the EDQNM model for $E(k)$
whose dynamical equation is only (\ref{eqLin}).

The corresponding set of time-dependent differential equations are advanced in 
time starting from initial conditions which correspond to a narrow-band distribution
of kinetic energy centered about a prescribed peak. The spectra then evolve into
developing an inertial subrange and a dissipative one. From this time on,
turbulence evolves in a self-similar decay. 

The evolved rotating turbulence spectrum presented on figure  \ref{fig5} corresponds to
a Reynolds number $\textit{Re}=7000$ and the Rossby number is $\textit{Ro}=u/(\Omega L)=0.04$ ($u$
is the \textit{rms} velocity, $L$ the integral scale and $\Omega$ the rotation rate)
although the figures at initial time were higher, since they usually decay in time.
For the stratified turbulence spectrum, $\textit{Re}=1000$ and the Froude number
is $\textit{Fr}=u/(NL)=0.09$, where $N$ is the Brunt-V\"ais\"al\"a frequency which characterizes
the intensity of the vertical density gradient. Note that the Reynolds number is lower
in the stratified simulation, since a part of the dynamics is contained in the 
potential energy stored by density fluctuations around the mean density.

\subsection{Quick comments on the spectral statistics and the dynamics of anisotropic turbulence}
\label{edqnmres}
\begin{figure}
\centering
\unitlength 1mm
\begin{picture}(50,70)
\put(-15,5){\includegraphics[width=0.5\textwidth]{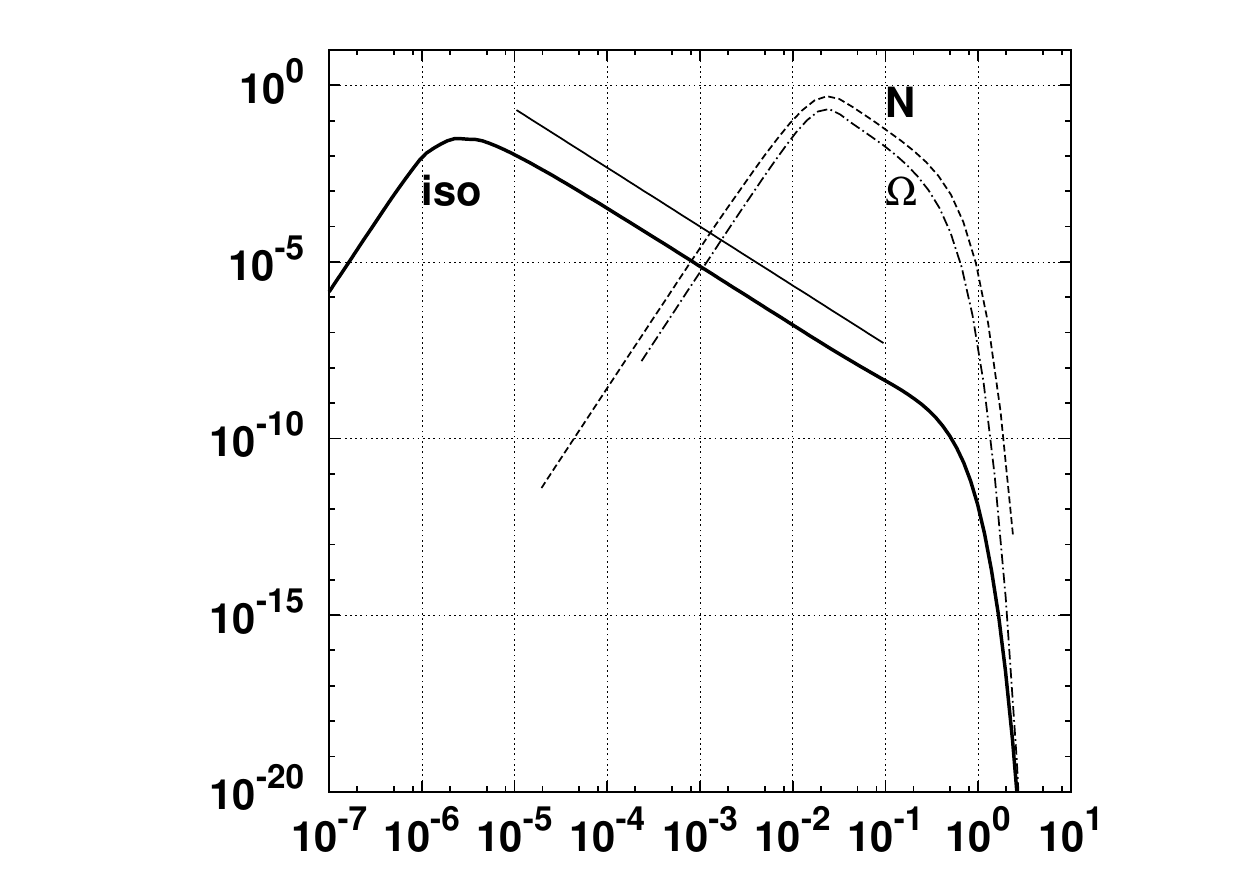}}
\put(50,0){$k\eta$}
\put(-5,33){\rotatebox{90}{$E(k)$}}
\end{picture}  
\caption{\label{fig5} Solid line: Kinetic energy spectrum $E(k)$ of isotropic turbulence
at $\textit{Re}^L=26\times 10^6$ given by the EDQNM closure model. Dot-dashed line: spherically accumulated
kinetic energy spectrum given by
the EDQNM$_2$ model for rotating turbulence. Dashed line: spectrum given
by the closure for stratified turbulence. The inertial range $k^{-5/3}$ power-law is also
shown on the figure, corresponding to the scaling of the isotropic spectrum. The spectra for rotating and stratified turbulence both scale like $k^{-1.9}$. The wavenumber $k$ is normalized  by the inverse of the Kolmogorov
length scale $\eta$.}
\end{figure}

The developed spectra for isotropic, rotating, and stably stratified turbulence are
plotted on figure  \ref{fig5}. Note that spectra of anisotropic turbulence, 
in our model, but also in DNS  or in experimental measurements, depend on 
the orientation of the wavevector, in agreement with two-point correlation
functions which depend on the orientation of the separation vector. The relevant
parameter is the angle $\theta$ with the axis of symmetry.

We have purposefully chosen an very high Reynolds number for the
isotropic spectrum for three reasons: (a) it demonstrates the capacity
of statistical EDQNM closures to reach Reynolds numbers closer to
those
observed in geophysical or astrophysical contexts than is possible with DNS; (b) the anisotropic
EDQNM$_2$ closure requires additional computational effort, and
therefore the inertial ranges observed on the corresponding 
spectra of figure  \ref{fig5} span roughly two decades and are reduced
with respect to the isotropic case; however, equivalent
DNS would require an order of magnitude larger computational effort; (c)
although the isotropic EDQNM spectrum exhibits six decades of
inertial subrange,
the corresponding kinetic energy transfer plotted on figure \ref{fig7}
only exhibits a narrow plateau at zero (if at all) between the energy-losing large scales
and the energy-gaining small dissipative scales, also corresponding to 
constant downscale spectral energy flux.

One should also note that statistical two-point models  are by essence
better adapted to produce smooth spectral statistics than Direct Numerical Simulations.
First, the discretization of the spectral space can be adapted easily when
solving EDQNM equations, by adjusting both the wavenumber $k$ distribution
and its orientations, with the possibility of accumulating grid points towards
the equatorial or the polar directions. By construction, DNS is constrained to 
a Cartesian uniform grid discretization of Fourier components. 
Moreover, the recent experiment by Lamriben \textit{et al.} \cite{lamriben2011}
has shown that ensemble averaging requested to compute smooth third-order statistics
--- namely third-order structure functions or energy transfers ---
required hundreds of realizations, an issue which also applies to DNS realizations.
Since the statistical two-point model equations already concern the statistics
of the second- and third-order moments, explicit ensemble averaging is 
self-contained in the model (as observed in the following smooth figures of results produced
by EDQNM).

The structure of decaying rotating turbulence has been described extensively from
experimental \cite{MOISY}, theoretical,  and numerical results 
 \cite{cambon97,yoshimatsu_columnar_2011}.
Inertial wave turbulence was studied by Galtier \cite{galtierwave03}, and  
Mininni \& Pouquet investigated energy and helicity spectra of rotating turbulence \cite{mininni09}, with different spectral scaling for each. The dynamics and
modified energy cascade produce an anisotropic structure which was
simulated with high resolution DNS also by Morinishi \textit{et al.}
 \cite{morinishi_dynamics_2001}. Anomalous scaling of structure functions
in rotating turbulence also permitted Seiwert \textit{et al.} to investigate
the intermittency of rotating turbulence \cite{seiwert}.

 The
interpretation of the dynamics leading to rotating homogeneous
turbulence structuration is still debated; 
the two main viewpoints rely either on linear timescales of inertial waves, with the Coriolis
force acting on locally inhomogeneous structures that emit waves \cite{staplehurst08}, or
on a long-term nonlinear effect \cite{cambon97}. It is not the object of this paper to
reconcile these viewpoints, but, in short, both agree that in rotating turbulence 
vortices are elongated along the axis of rotation. 
 The corresponding dynamics is that of preferential energy transfer towards
motion close to two-dimensional, although complete two-dimensionalization is not expected.
It also means that  energy accumulates in orientation dependent kinetic energy
spectra $E(k,\theta)$ in the vicinity of the two-dimensional manifold $\theta\simeq\pi/2$ which is the neighborhood
of horizontal wavevectors in spectral space. This is confirmed by the angular
dependent energy spectrum plotted on the top panel of figure  \ref{fig6}, which also compares well
with spectra processed from   DNS of decaying rotating turbulence. Note that
the power-law scaling of the corresponding spectrum in figure  \ref{fig5} is the
result of the spherical averaging over the directional spectra of figure   \ref{fig6}, with
no single identical power-law applying to each individual one. 
The bottom panel of figure \ref{fig6} presents a tentative measure of the anisotropy
in the spectra, by subtracting to the directional spectral the averaged spectrum. The curves
present only the limit spectra, equatorial or polar; also note that the anisotropy is \textit{reversed} between the rotating and the stratified cases, with more energy in the polar
direction (along the symmetry axis) in stratified turbulence, whereas the accumulation
of spectral energy in rotating turbulence is in the vicinity of the equatorial (perpendicular
to the symmetry axis) direction.

For the stably stratified case, the spectral structuration is reversed, such that
the energy accumulates in spectra of vertical  $\bm{k}$ ($\theta\simeq 0$) as shown on figure
 \ref{fig6} (see also \textit{e.g.} Lindborg \cite{lindborg06}). Again, the anisotropic statistical model compares extremely well
with experimental wind-tunnel observations and DNS \cite{godeferd2003}. These
comparisons not only concern dynamical quantities, such as energy or dissipation,
but also directional integral length scales, and the most refined possible comparison
between directional spectra $E(k,\theta)$, that is for every direction of every scale. In both the stratified
and the rotating cases, EDQNM$_2$ model predictions are very close to the
same spectral statistics extracted from DNS fields, down to the most refined
ones.
In physical space, the structuration of stably stratified turbulence
 corresponds to a layering of the flow, with 
strong horizontal motion, and  vanishing vertical velocity, although 
the flow retains a strong vertical \textit{variability} due to the presence
of large vertical gradients $\partial/\partial z$.  \cite{yoon_evolution_1990,praud_decaying_2005}

The anisotropy observed in the spectra of figure  \ref{fig6} is the departure
of each spectrum $E(k,\theta)$ around the average shown on figure  \ref{fig5}. 
Since the plot is in logscale, the clear difference between horizontal
wavenumber spectra and vertical ones demonstrates the large ratio between the
energy in vertical and horizontal motion. Moreover, one observes that in the
stratified turbulence spectrum the anisotropy is largest at the larger, energy-containing,
scales, although it extends to the middle of the inertial sub-range;  in the 
rotating turbulence spectrum, the anisotropy extends throughout the inertial
range down to the smallest scales. Of course, this observation  depends
on the values of the Rossby and Froude numbers as well as on the Reynolds number.
The rotation (or stratification) timescale has to be compared to the
local timescale of a given turbulent structure associated with wavenumber
$k$ in order to assess whether its dynamics may be affected by the Coriolis
or the Boussinesq forces (for instance by computing a Zeman scale $(\Omega^3/\epsilon)^{-1/2}$ \cite{MININNI-ROSENBERG-POUQUET}, where $\epsilon$ is the dissipation, or similarly
an Ozmidov scale
for stratified turbulence by replacing $\Omega$ by $N$ \cite{Ozmidov}).
It nonetheless demonstrates that there exist parameter ranges at which turbulence
may be strongly affected by external distortions, and its structure and dynamics
can depart significantly from that of isotropic turbulence.

In addition to providing   the second-order statistics corresponding to the two-point
energy spectra discussed above, the closure provides quantitative
information on third-order two-point correlation spectra, \textit{i.e.} the
kinetic energy transfer spectra shown
on figure  \ref{fig7}. Apart from the Reynolds number difference,
the spherically averaged transfer spectra of anisotropic turbulence are
similar to the isotropic turbulence transfer. Note that we present 
spectra of the kinetic energy transfer $T^e(k)$, which is only a part
of the energy transfers occurring in stratified turbulence, since there is
also coupling with the potential energy mode. As observed in  the isotropic
case, in the anisotropic cases energy is drawn from the large scales and injected in
 the small scales in
a classical forward cascade. In the mean time, if one observes the
orientation-dependent transfer spectra (not presented here), one also notices  a re-distribution
of energy among different orientations of wavevectors, even at constant
wavenumber,    \textit{i.e.} without interscale exchange.

\begin{figure}
\centering
\unitlength 1mm
\begin{picture}(50,70)
\put(-15,5){\includegraphics[width=0.5\textwidth]{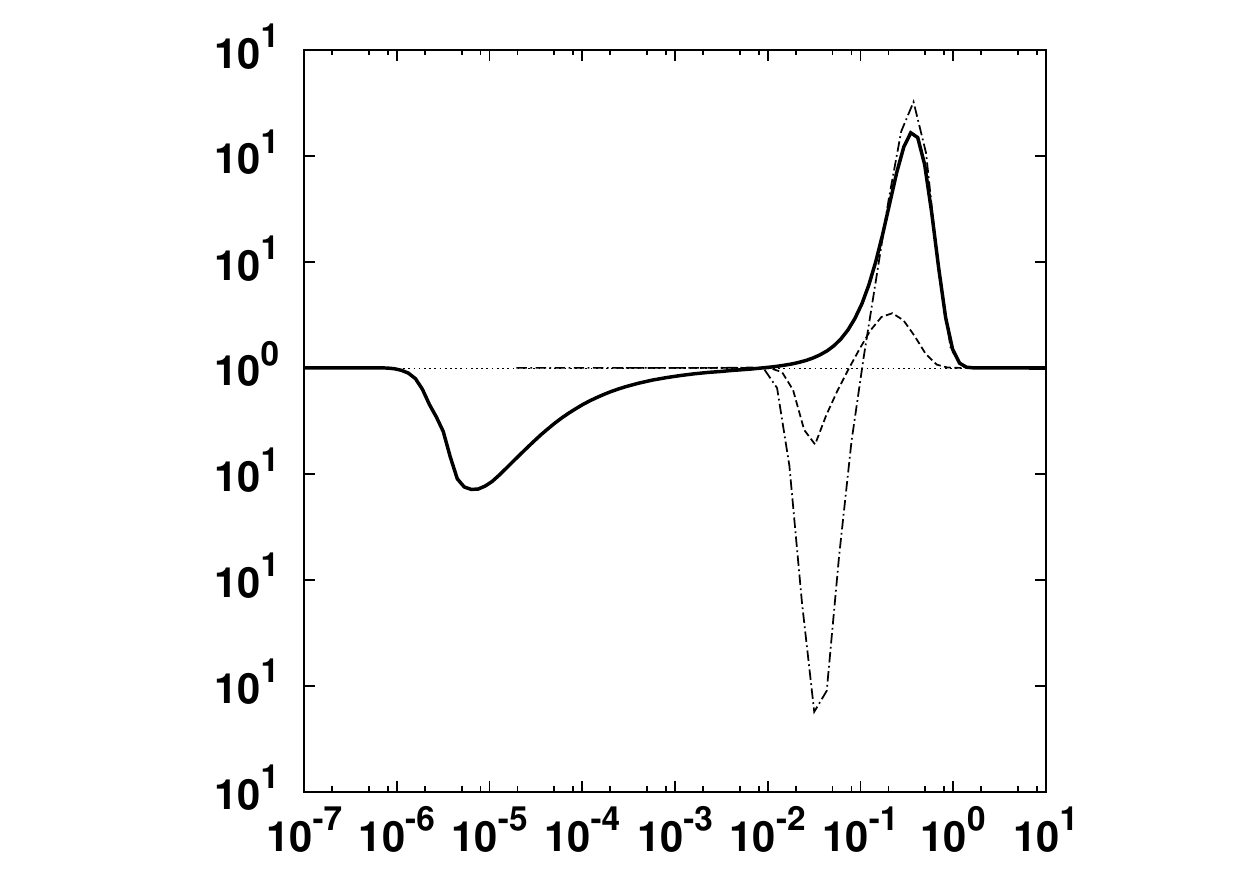}}
\put(50,0){$k\eta$}
\put(-5,33){\rotatebox{90}{$T(k)$, $T^e(k)$}}
\end{picture}  
\caption{\label{fig7} Solid line: Kinetic energy transfer spectrum $T(k)$ of isotropic turbulence
given by the EDQNM closure model (scaled by $10^2$ to render it visible). Dot-dashed line: spherically accumulated
kinetic energy transfer spectrum $T^e(k)$  given by
the EDQNM$_2$ model for rotating turbulence. Dashed line: transfer spectrum given
by the closure for stratified turbulence. }
\end{figure}
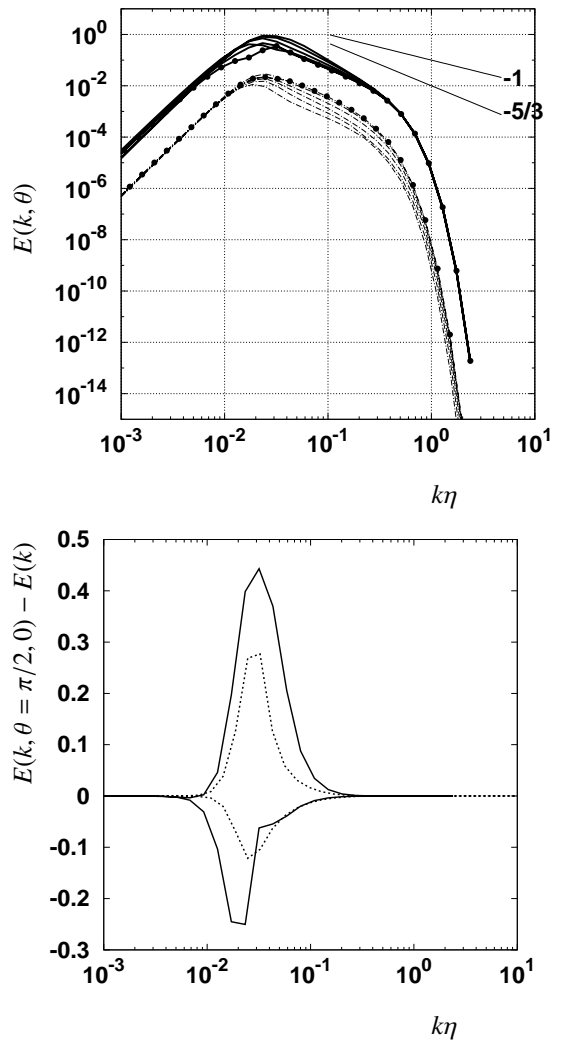
\begin{figure}
\centering
\unitlength 1mm
\begin{picture}(50,70)
\put(-15,5){\includegraphics[width=0.5\textwidth]{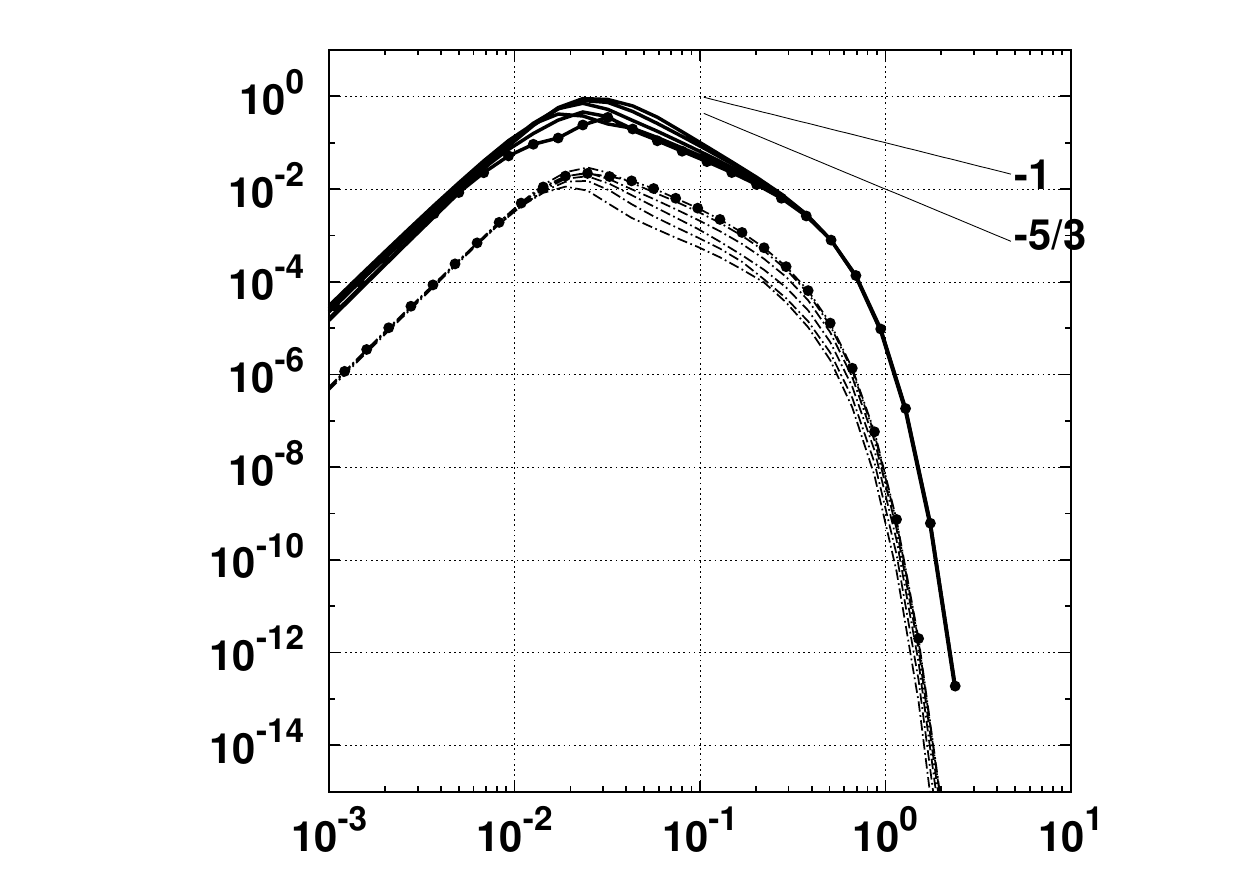}}
\put(50,0){$k\eta$}
\put(-5,33){\rotatebox{90}{$E(k,\theta)$}}
\end{picture}  

\centering
\unitlength 1mm
\begin{picture}(50,70)
\put(-15,5){\includegraphics[width=0.5\textwidth]{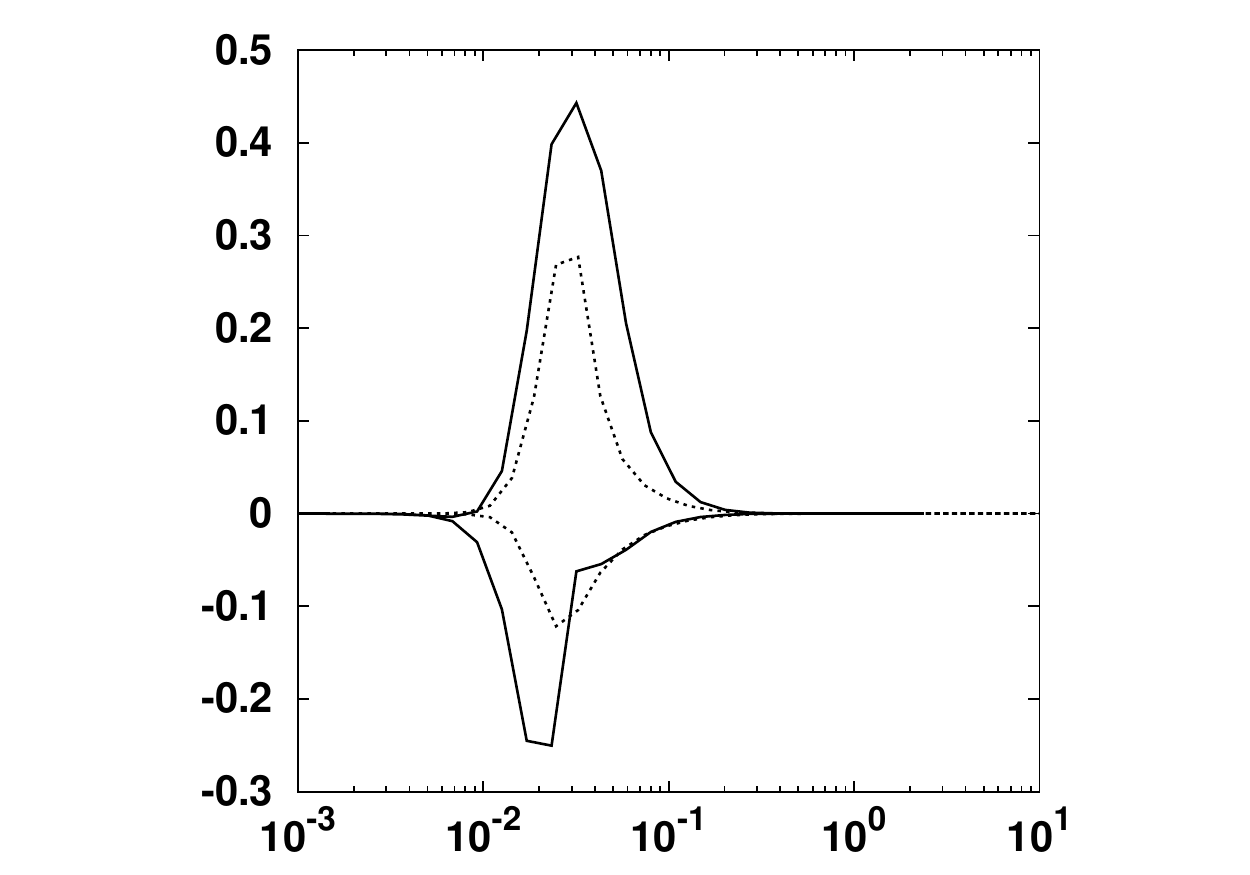}}
\put(50,0){$k\eta$}
\put(-5,33){\rotatebox{90}{$E(k,\theta=\pi/2,0)-E(k)$}}
\end{picture}  

\caption{\label{fig6} Top panel: Orientation-dependent kinetic energy spectra $E(k,\theta)$  (
seen to be monotonically varying with $\theta$). Solid
lines: results from  EDQNM$_2$ model for stratified turbulence; dashes: results from
the  EDQNM$_2$ model for rotating turbulence, shifted down one decade. For each case, 
a subset of the discretized orientations $\theta$ are presented.  Spectra
corresponding to horizontal wavevectors are indicated by black dots.
Bottom panel: A measure of the anisotropy showing the  difference between the polar and equatorial spectra with the spherically averaged one. Solid lines: EDQNM$_2$ result for stratified turbulence, dashed lines for the rotating results.  
}
\end{figure}
%


\subsection{Structure functions}
\label{physres}
The above observations of second- and third-order correlations spectra
have a direct impact on the second- and third-order structure functions
of turbulence, according to the duality of spectral and physical space.
Some passage relations between the two spaces have been 
presented in section  \ref{sec1}. We now present the results obtained by using
these relations applied to the EDQNM closure results for isotropic and
anisotropic turbulence. It is however not possible to derive any $n^{th}$-order
structure function from spectral information contained in two-point statistical
closures, precisely because the spectral information is restricted to
$n^{th}$-order correlations at  \textit{two} points. A case-by-case inspection is thus required.

\begin{figure}
\centering
\unitlength 1mm
\begin{picture}(50,70)
\put(-15,5){\includegraphics[width=0.5\textwidth]{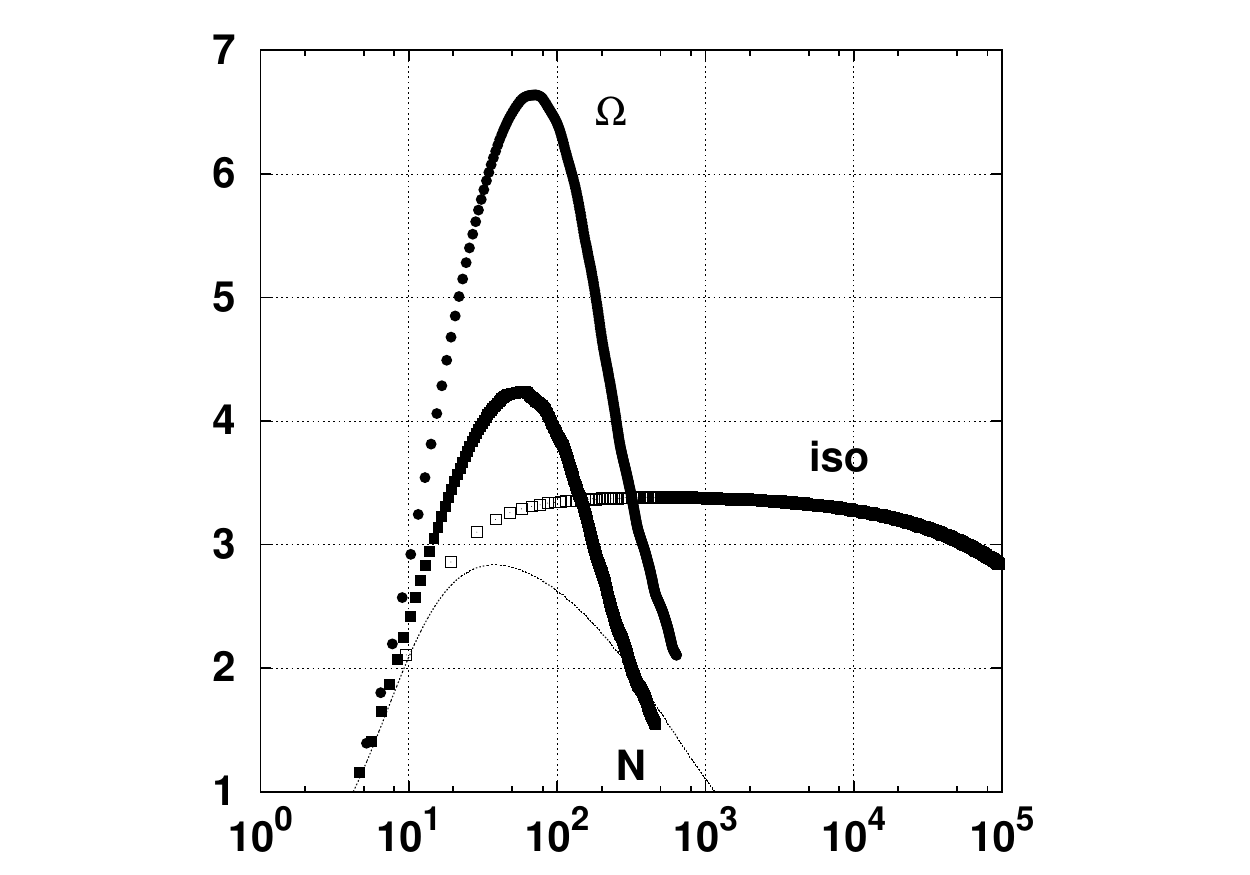}} 
\put(50,0){$r/\eta$}
\put(-5,33){\rotatebox{90}{$\langle (\delta u_i\delta u_i)\rangle / (\varepsilon r)^{2/3}$}}
\end{picture}  
\caption{\label{fig1} Second-order structure function $\langle(\delta u_i\delta u_i)\rangle$ computed from (\ref{eqqdeux}) normalized
by $(\varepsilon r)^{2/3}$ as a function of the separation $r$ normalized by the Kolmogorov 
length scale $\eta$. 
Open squares: results from the EDQNM model for isotropic turbulence; 
Filled squares: results from EDQNM$_2$ model for stratified turbulence;
Filled circles: results from EDQNM$_2$ model for rotating turbulence.
The dotted line presents isotropic turbulence EDQNM results at $\textit{Re}=2600$.} 
\end{figure}

The relation (\ref{eqqdeux}) linking the kinetic energy spectrum to the
second-order structure function is applied to the results obtained by EDQNM,
and also to the anisotropic turbulence results. 
The numerical resolution of the isotropic EDQNM model directly provides
discretized distribution of $E(k)$ and $T(k)$. For the EDQNM$_2$, the spherically
averaged spectra $E(k)$ and $T(k)$ are obtained from the anisotropic 
spectral data $E(k,\theta)$ and $T^e(k,\theta)$, by averaging over $\theta$.
Once these data are obtained, a simple numerical quadrature of the integrals in
equations (\ref{eqqdeux}) and (\ref{thstr}) is used to obtain $\langle\left( \delta u_i  \delta u_i \right) \rangle$ and $\langle\delta u\left(\delta q\right)^2\rangle$ (likewise for $\langle(\delta u)^3\rangle$ with an integral not recalled here).
The corresponding structure
function is normalized by $(\varepsilon r)^{2/3}$  and plotted against
 $r/\eta$ ($\eta$ is the Kolmogorov length scale) on figure  \ref{fig1},  showing the 
scaling which is obtained for isotropic turbulence, and illustrating the advantage
brought by achieving a high value of the Reynolds number. For rotating turbulence and
for stably stratified turbulence, the corresponding values of the Reynolds number
are lower, and one has to deal with a narrower curve. It is clear, however, that the
isotropic scaling is far from being applicable to the anisotropic turbulence structure functions,
since, here, if we compute $\textrm{max}(\langle \left( \delta u_i  \delta u_i \right) \rangle/(\varepsilon r)^{2/3})$, one obtains 3.4 for the isotropic case, 4.2 for the stratified
case, 6.6 for the rotating case. 

\begin{figure}
\centering
\unitlength 1mm
\begin{picture}(50,70)
\put(-15,5){\includegraphics[width=0.5\textwidth]{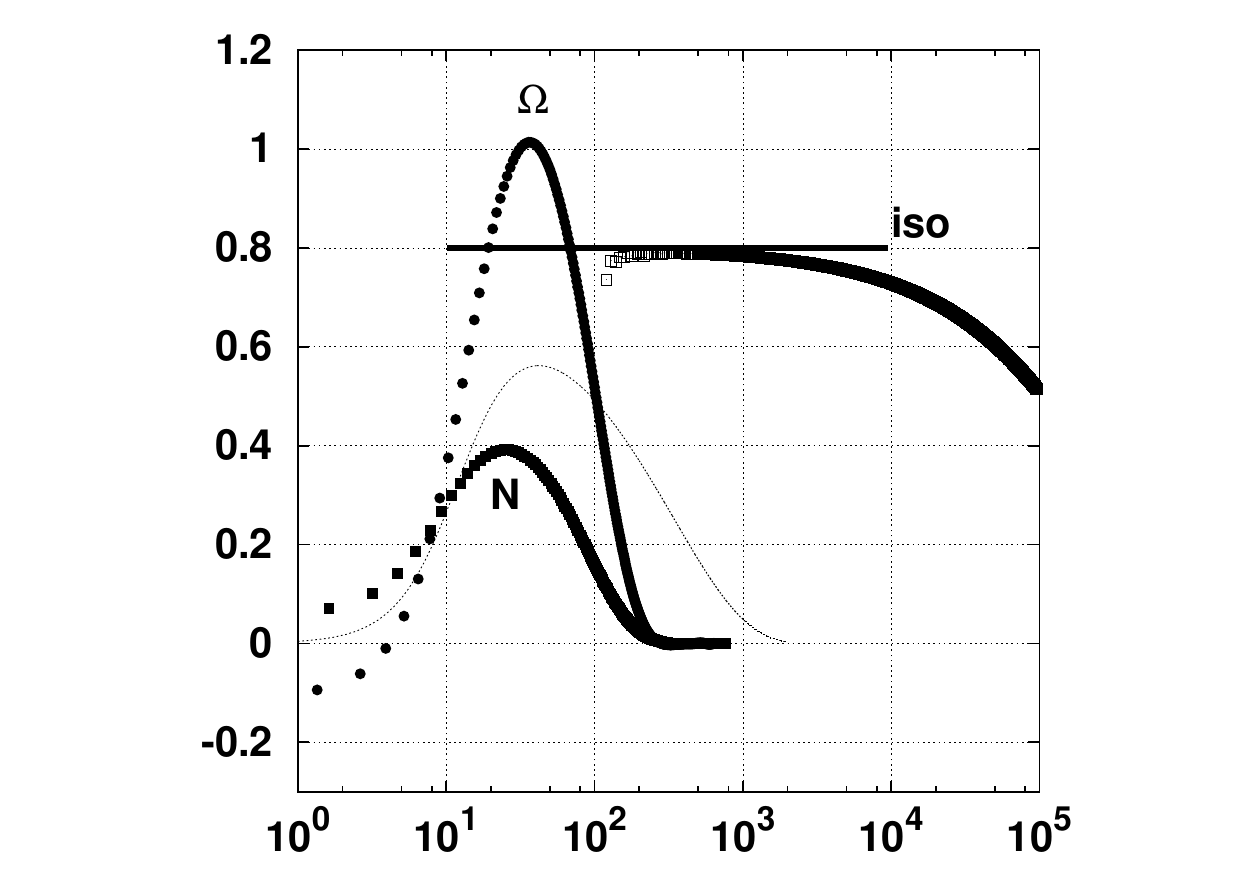}}
\put(50,0){$r/\eta$}
\put(-5,33){\rotatebox{90}{$-\langle(\delta u)^3\rangle/(\varepsilon r)$}}
\end{picture}  
\caption{\label{fig2} Third-order structure function of the longitudinal
velocity increment $\langle(\delta u)^3\rangle$ normalized by $-(\varepsilon r)$ as a function
of the separation normalized by the Kolmogorov length scale. Same symbol convention as 
in figure \ref{fig1}. The horizontal line exhibits the $4/5$ scaling. The dotted line presents isotropic turbulence EDQNM results at $\textit{Re}=2600$.}
\end{figure}

\begin{figure}
\centering
\unitlength 1mm
\begin{picture}(50,70)
\put(-15,5){\includegraphics[width=0.5\textwidth]{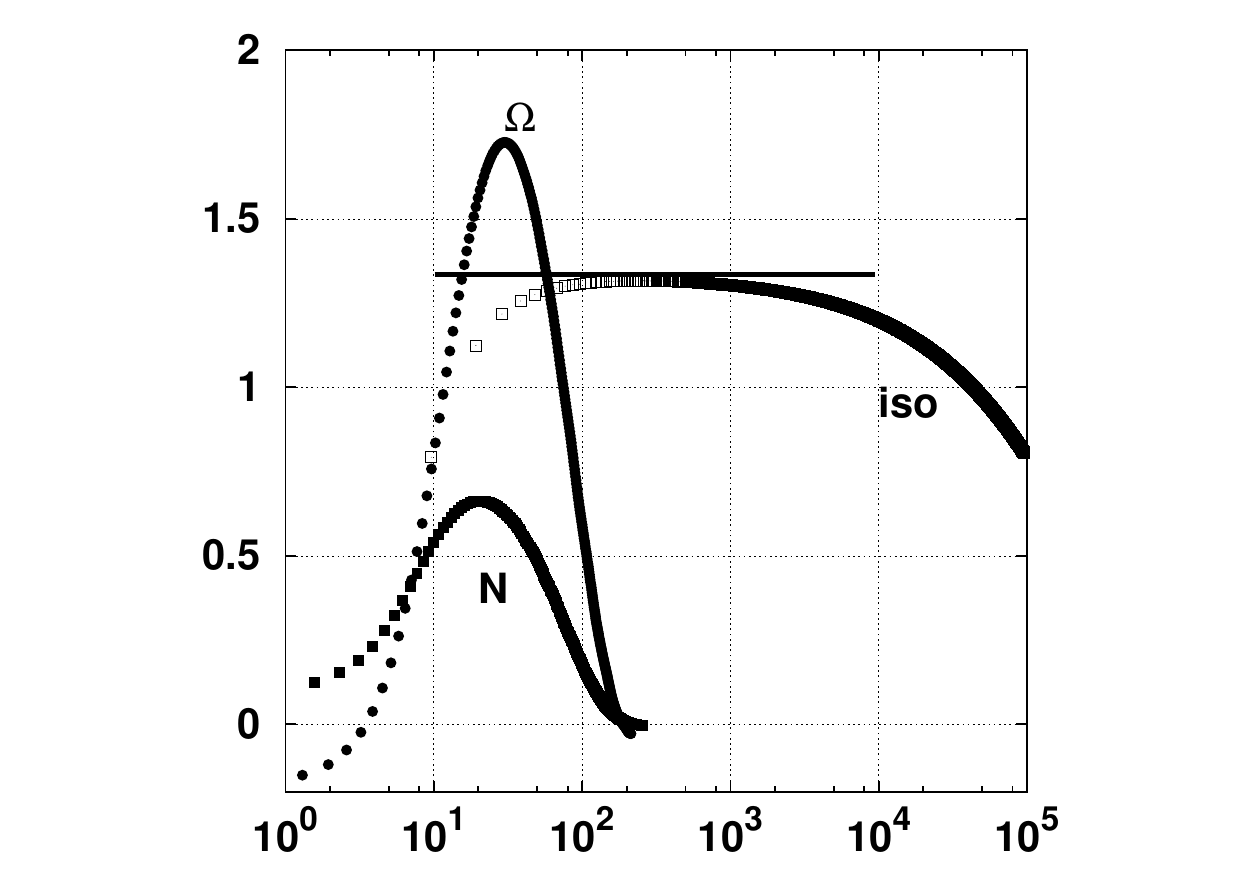}}
\put(50,0){$r/\eta$}
\put(-5,33){\rotatebox{90}{$-\langle\delta u (\delta u_j\delta u_j)\rangle/(\varepsilon r)$}}
\end{picture}  
\caption{\label{fig3} Third-order structure function $\langle\delta u (\delta u_j\delta u_j)\rangle$
as proposed by \cite{antonia97} normalized by $-(\varepsilon r)$ as a function
of the separation normalized by the Kolmogorov length scale. Same symbol convention as 
in figure \ref{fig1}. The horizontal line exhibits the $4/3$ scaling.}
\end{figure}

This is also observed on the curves for the third-order structure functions
plotted on figures  \ref{fig2} and  \ref{fig3} in two different forms. Figure   \ref{fig2}
shows the averaged cubed longitudinal increment $\langle(\delta u)^3\rangle$ divided
by the behavior proposed in Kolmogorov theory: $-(\varepsilon r)$.  The expected proportionality
constant $4/5$ is clearly  recovered by the EDQNM closure transfer processed through
equation (\ref{eqdix}), with a nice plateau tangential to the $y=4/5$ horizontal line. 
It should be noted that even at this very high value of the Reynolds number the
extent of the plateau is less than a decade in scale. This may explain part of the
early controversies about the precise value of the proportionality constant obtained
from experimental measurements. The same considerations apply for the other form of the 
third-order structure function proposed by \cite{antonia97}, $\langle\delta u (\delta u_j\delta u_j)\rangle$, with a proportionality constant $4/3$ with $\varepsilon r$, plotted
on figure  \ref{fig3} using equation (\ref{anton}). In both plots, the scaling is obviously
not the right one for
rotating turbulence or for stratified turbulence, although the order
of magnitude of the peaks of the structure functions for the anisotropic
turbulence cases are consistent with the isotropic
turbulence one. ($\textrm{max}(\langle\delta u (\delta u_j\delta u_j)\rangle)/(\varepsilon r))$
is marginally lower than 0.8 in the isotropic case, 0.39 for stratified results and 1.02 for 
the rotating results.)

 In rotating turbulence, the energy cascade is known to be reduced 
by the presence of rotation, so that the dissipation is lower than for an equivalent
Reynolds number isotropic turbulence dynamics. This can explain the fact that
the peaks of $-\langle\delta u (\delta u_j\delta u_j)\rangle/(\varepsilon r)$ or of 
$-\langle(\delta u)^3\rangle/(\varepsilon r)$ are higher than in isotropic turbulence.
In addition, one observes on figures \ref{fig2} and \ref{fig3} a negative excursion of the third-order
structure function in the very large scales, which could be attributed to a trend towards a
reverse cascade, although this has to be confirmed by other simulations.

In the stratified case, the trend is reversed, an effect which could be attributed
to augmented dissipation. The analysis is subtle here, since there is not only a
cascade of energy due to classical nonlinear turbulent interaction, but there is also
a coupling between the kinetic energy and the potential energy. It seems that the
balance between energy taken away from the velocity field for creating density
fluctuations is more than compensated by the  increased dissipation from the
added scalar cascade. An attempt at including the part of dissipation arising
in the potential energy balance fails at correcting enough the scaling of the
third-order structure functions of stratified turbulence in figures  \ref{fig2} and  \ref{fig3}.

Finally, in order to address the Reynolds number dependence issue for
the third-order structure function convergence, we computed an isotropic turbulence
EDQNM result for a lower Reynolds number than proposed above, at $\textit{Re}=2600$,
of the same order of magnitude as those for the anisotropic EDQNM$_2$ data.
The solid curve of figure \ref{fig2} shows the corresponding result, which underestimates
significantly the 4/5 constant. The departure amplitude is similar to that of the anisotropic
data, although we believe it is of different nature. Following the analysis by Taylor \textit{et al.} \cite{taylorkurien}, there are therefore two sources of departure to the K41 scaling, one
due to low Reynolds number effect, the other to anisotropy. Here, for the anisotropic
runs, anisotropy is responsible for the departure from the asymptotic large Reynolds number
scaling, since for rotating turbulence it leads to an \textit{overestimate}, whereas for
stratified turbulence it is \textit{underestimated}. Also, the measures presented by
Antonia \& Burattini for isotropic turbulence exhibit a uniform convergence from \textit{below} the value 4/5 \cite{antonia06}. Thus, although there may be a finite
Reynolds number effect, we believe that, to first order, the difference with the
high Reynolds number isotropic scaling is due to anisotropy. This argument is 
also supported by the curve for the second-order structure function in isotropic turbulence
at the lower Reynolds number $\textit{Re}=2600$ presented on figure~\ref{fig1}. It
shows that although the asymptotic large Reynolds number 
scaling of  is not yet reached, the maximum of 
$\langle (\delta u_i\delta u_i)\rangle / (\varepsilon r)^{2/3}$ is within 30\%
of the asymptotic value, whereas the peak for rotating turbulence is twice
this value. The shape of the spectra and the difference in the downscale cascading rate
in the anisotropic runs with respect to the isotropic one can only explain this 
departure. 

\section{Conclusion and perspectives}

We have described in this work the spectral statistics of anisotropic turbulence,
with spectra of second- and third-order two-point  correlations. These kinetic
energy spectra and kinetic energy transfer spectra are the result of the spherical
averaging of spectra of correlations taken with separations 
along different angles with respect to the axis of symmetry, or in other words
for different orientations of the wavevectors. Not only these angle-dependent
spectra correspond to variable intensities of the energy, but their inertial range
scaling, when averaged, produces the power-law scaling of the overall kinetic energy spectrum.
Results obtained by a two-point statistical closure of EDQNM type show that,
at high Reynolds number, the resulting anisotropy can be strong, and corresponds
to a non classical structuration of the flow. The cases of strongly rotating turbulence
and strongly stratified turbulence were investigated, with respectively an elongation
of the vortices along the rotation axis or a horizontal layering. 
The velocity components are consequently of variable magnitude in the vertical
and horizontal directions, so that an impact on the velocity second-order structure functions
is expected. 
The access to the detailed energy transfers also permits the computation
of third-order structure functions, and the comparison with exact
scalings obtained for isotropic turbulence. As expected, the constants
obtained are different, although thanks to the relationships presented
in section  \ref{sec1} one is able to closely related the anisotropic
spectral characterization to the structure functions, thus to link the dynamics
of energy cascade and inter-orientation transfer to the statistical
moments in physical space. 
The departure from K41 theory for anisotropic
turbulence statistics of the second- and third-order structure
functions shows that 
the K\`arm\`an-Howarth equation from which theoretical scalings are derived
exhibits a modified equilibrium when the structure of turbulence is anisotropic.
The correspondence with the Lin equation shows that the dynamics of anisotropic
turbulence is also modified, and we are able to trace the origin of the
modified energy spectra scalings to the transfer terms. Since these transfer
terms are accessible by spectral theory, this may be a way to improve
the prediction of the third-order structure function moment in rotating or 
stratified turbulence, using two-point statistical closures.

Of course, this  calls for additional work and further developments, including the derivation
of  relationships equivalent to  (\ref{eqqdeux}) and (\ref{anton}) but considering
the specific components of velocity parallel or orthogonal to the axis of symmetry.
In so doing, one will highlight the value of the anisotropic EDQNM$_2$ closure which
not only allows to model high Reynolds number turbulence but also provides
quantitative information on  the anisotropy of statistics.

\medskip
This work is funded by the French \textit{Agence Nationale de la Recherche} under grant number 
ANISO-340803. The reviewing work and suggestions of the Referees is also thankfully acknowledged.

%


\bibliographystyle{elsarticle_num}
\bibliography{phys.bib}


\end{document}